\documentclass[reprint,amsmath,amsfonts,prl,aps,superscriptaddress]{revtex4-1}
\usepackage{graphicx} 
\usepackage{braket} 

\newcommand{\op}[1]{\mathcal{#1}} 
\newcommand{\PT}{\op{PT}} 

\newcommand{\E}{\mathrm{e}} 
\newcommand{\I}{\mathrm{i}} 

\newcommand{\hc}{\text{h.c.}} 
\newcommand{\diag}{\mathrm{diag}} 

\newcommand{\va}{v} 
\newcommand{\vb}{w} 
\newcommand{\vt}{v_T} 

\begin{document}
\title{Fragile aspects of topological transition in lossy and parity-time symmetric quantum walks}

\author{Andrew K. Harter}
\affiliation{Department of Physics, Indiana University-Purdue University Indianapolis (IUPUI), Indianapolis, Indiana 46202, USA}
\author{Avadh Saxena}
\affiliation{Theoretical Division and Center for Nonlinear Studies, Los Alamos National Laboratory, Los Alamos, New Mexico 87545, USA}
\author{Yogesh N. Joglekar}
\affiliation{Department of Physics, Indiana University-Purdue University Indianapolis (IUPUI), Indianapolis, Indiana 46202, USA}

\date{\today}
\begin{abstract}

Quantum walks often provide telling insights about the structure of the system on which they are performed. In $\mathcal{PT}$-symmetric and lossy dimer lattices, the topological properties of the band structure manifest themselves in the quantization of the mean displacement of such a walker. We investigate the fragile aspects of a topological transition in these two dimer models. We find that the transition is sensitive to the initial state of the walker on the Bloch sphere, and the resultant mean displacement has a robust topological component and a quasiclassical component. In $\mathcal{PT}$ symmetric dimer lattices, we also show that the transition is smeared by nonlinear effects that become important in the $\mathcal{PT}$-symmetry broken region. By carrying out consistency checks via analytical calculations, tight-binding results, and beam-propagation-method simulations, we show that our predictions are easily testable in today's experimental systems.

\end{abstract}

\maketitle

\section{Introduction}
\label{sec:introduction}

Non-Hermitian systems that are invariant under combined operations of parity ($\mathcal{P}$) and time-reversal ($\mathcal{T}$) symmetry have been intensely explored over the past five years \cite{El-Ganainy2018}. Such continuum, $\mathcal{PT}$-symmetric Hamiltonians have a region in their parameter space where the eigenvalues are entirely real~\cite{Bender1998,Bender2002} and the eigenfunctions are simultaneous eigenfunctions of the antilinear $\PT$ operator (the $\PT$-symmetric phase). When the strength of the nonhermiticity is increased, the spectrum changes to complex-conjugate pairs ($\PT$-symmetry broken phase). In the past decade, it has become clear that, while not fundamental in nature \cite{Lee2014}, $\mathcal{PT}$ Hamiltonians faithfully describe open classical systems with balanced, spatially separated loss and gain \cite{Klaiman2008,El-Ganainy2007,Joglekar2013}. The dynamics that result from such Hamiltonians have been experimentally observed in a wide variety of setups including coupled optical waveguides \cite{Ruter2010}, optical resonators \cite{Regensburger2012,Peng2014}, silicon photonic circuits~\cite{Feng2011}, metamaterials~\cite{Feng2012,Lin2011}, and microring lasers \cite{Hodaei2014,Feng2014}. 

Interest in the topological properties of lattice models can be traced back to the studies of polyacetylene chain excitations by Su, Schrieffer and Heeger (SSH) \cite{Su1979,Heeger1988}. The SSH model is a one-dimensional chain of strongly-bonded pairs of atoms (dimers) that in turn are connected by weak bonds. It has topologically protected edge states and shows a variety of robust phenomena including the existence of fractional-charge excitations~\cite{Heeger1988}. There have been extensive studies of its variations, such as the SSH model with site-dependent periodic potential \cite{Li2014}, Aubry-Andre Harper (AAH) models \cite{Harper1955,Aubry1980} where the unit cell comprises $p\geq 3$ atoms~\cite{Ganeshan2013,Verbin2013,Schomerus2013}, and quasiperiodic models \cite{Kraus2012}. Topological properties of the SSH model also manifest themselves in its lossy counterpart. Early theoretical work by Rudner and Levitov~\cite{Rudner2009} (referred to as RL) showed that in an SSH model with absorption on alternate sites, a quantum walker starting on a neutral site has a quantized mean displacement $\langle\Delta m\rangle_L$ before it is absorbed. When the coupling $v$ within a dimer is much larger than the inter-dimer coupling $w$, the walker is absorbed within the same dimer giving $\langle\Delta m\rangle_L=0$. In the other limit, $w\gg v$, the walker is far more likely to reach the loss-site in the adjacent dimer than the loss-site within its initial dimer, giving $\langle\Delta m\rangle_L=1$. What is surprising is that the mean displacement remains quantized at zero when $v\gtrsim w$ and at unity when $v\lesssim w$. RL showed that the quantization of $\langle \Delta m\rangle_L$ is due to its equivalence with the topological winding number of the SSH band structure. This  topological transition in $\langle\Delta m\rangle_L$ was experimentally observed in a lossy waveguide array by Zeuner and coworkers~\cite{Zeuner2015} (referred to as Z), where its robustness against changes in couplings and loss rate was also demonstrated. 

Here, we consider a $\mathcal{PT}$-symmetric SSH model with gain and loss in each dimer. Apart from a shift along the imaginary axis, the Hamiltonian for a $\mathcal{PT}$ dimer is the same as the Hamiltonian for a dimer with one neutral and one lossy site \cite{Guo2009,Ornigotti2014}. Thus we expect that the $\PT$-symmetric SSH model will undergo two transitions, namely the $\PT$-symmetry breaking transition and a topological transition in the quantization of a {\it suitably defined mean-displacement} $\langle \Delta m\rangle$. In this paper, we investigate this topological transition and its fragile aspects. As we will show below, they are driven by the walker's initial state in a single dimer and the nonlinearities that are relevant in the $\mathcal{PT}$-broken phase of the SSH model. 

The plan of the paper is as follows. In the next section, we begin with an overview of the $\PT$-symmetric SSH model, its relation to the lossy SSH model, and a summary of results for $\langle\Delta m\rangle$. We examine the sensitivity of the mean-displacement $\langle\Delta m\rangle$ to the initial condition of the walker in Sec.~\ref{sec:initial-state}. We show that the mean displacement has a fragile quasiclassical component in addition to the robust topological component {\it a la} RL~\cite{Rudner2009}. This result is valid for both $\mathcal{PT}$-symmetric and lossy SSH models. In Sec.~\ref{sec:nonlinearity}, we present the results for changes in $\langle\Delta m\rangle$ due to the effects of nonlinear terms. We find that the quantization of the mean displacement $\langle\Delta m\rangle$ is destroyed in an antisymmetric manner across the topological transition threshold. We support our findings by analytical, tight-binding, and beam-propagation method (BPM) calculations. 


\section{$\mathcal{PT}$-symmetric SSH Model}
\label{sec:pt-dimer}

\begin{figure*}
\begin{center}
\includegraphics[width=\textwidth]{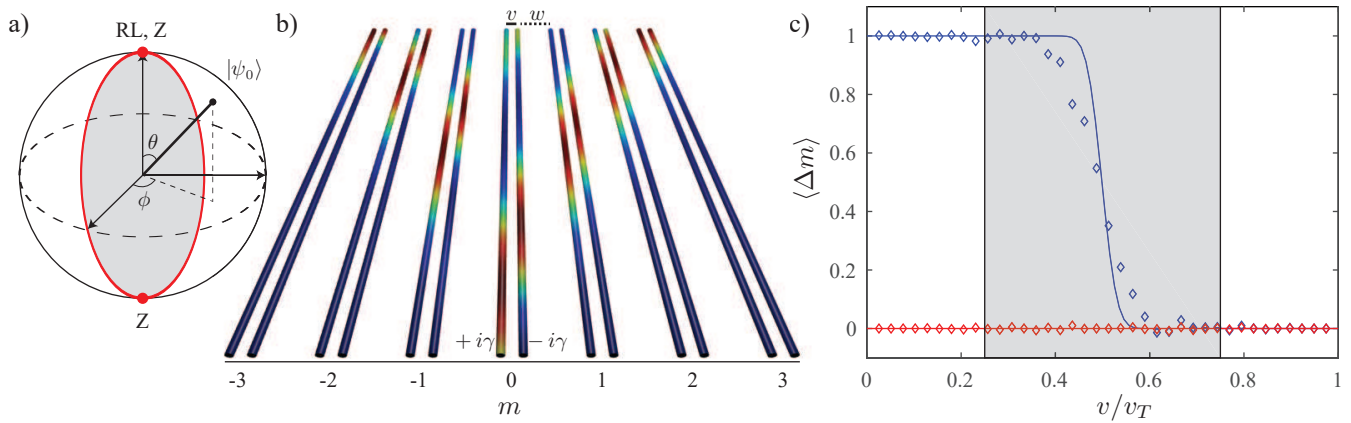}
 \caption{\label{fig:overview} (a) Bloch sphere of possible single-dimer localized initial states $\ket{\psi_0}=\cos(\theta/2)\ket{A} +e^{\I\phi}\sin(\theta/2)\ket{B}$. The north-pole and south-pole states were studied in a lossy SSH lattice by RL~\cite{Rudner2009} and Z~\cite{Zeuner2015}. (b) $\PT$-symmetric SSH waveguide array with intra-dimer coupling $\va$, inter-dimer coupling $\vb$, and gain-loss potentials $\pm i\gamma$ on sites A and B within each dimer. The color shows the site and time-dependent intensity obtained from a tight-binding calculation with initial state $|\Psi(0)\rangle=|0,A\rangle$. (c) The mean displacement $\braket{\Delta m}$, Eq.(\ref{eq:dm}) as a function of $\va/\vt$ 
for a $\PT$-symmetric dimer lattice. $\langle\Delta m\rangle$ shows a step-structure at $\va/\vt=1/2$ when $|\psi_0\rangle$ is at the north pole (blue) and is identically zero when $|\psi_0\rangle$ is at the south pole (red). Solid lines depict tight-binding results with $N=41$ dimers and $\gamma=0.5\vt$. The diamonds are from BPM simulations with $N=21$ waveguides~\cite{Zeuner2015,Harter2016b}. The coupling values $|\va-\vb|\leq\gamma$ for the $\PT$-symmetry broken phase, where nonlinear effects are dominant, are shown by the shaded window.}
\end{center}
\end{figure*}

Consider a lattice of $N=2M+1$ dimers (unit cells) labeled by integer $m\in[-M,M]$, where each identical dimer has two sites which we label $A$ and $B$. The coupling between $A$ and $B$ sites within each dimer is given by $\va$ and that between different dimers is $\vb$. The $A$ site within each dimer has an amplifying gain potential $+i\gamma$ and the $B$ site has a balanced loss potential $-i\gamma$. Using the basis kets $\ket{m,A}$ and $\ket{m,B}$, which correspond to states spatially localized to sites $A$ and $B$ respectively of the $m$th dimer, we describe the Hamiltonian for this system as $H_{PT}=H_0+\Gamma_{PT}$, where 
\begin{eqnarray}
H_{0} & = & \va\sum_{m=-M}^{M} \left(\ket{m,A}\bra{m,B} +\hc\right) \nonumber \\
& + & \vb\sum_{m=-M}^{M-1} \left(\ket{m,A}\bra{m+1,B} +\hc\right),\label{eq:hssh}\\
\Gamma_{PT} & = & \I\gamma\sum_{m=-M}^{M} \left(\ket{m,A}\bra{m,A} - \ket{m,B}\bra{m,B}\right).\label{eq:GammaPT}
\end{eqnarray}
Due to the translational invariance of this system under periodic boundary conditions, the Hamiltonian $H_{PT}$ is block diagonal in the momentum space, and reduces to a 2$\times$2 matrix for each momentum $k$, i.e. $H_{PT}(k) = \vec{h}(k)\cdot\vec{\sigma}$ where $k=2\pi n/N$ and $n=1,\ldots,N$,  $\vec{\sigma}=(\sigma_x,\sigma_y,\sigma_z)$ is the Pauli matrix vector, and the component functions are $h_x(k)\pm ih_y(k) = \va + \vb e^{\pm ik}$, and $h_z(k) = \I\gamma$. $H_{PT}(k)$ is a $\PT$-symmetric matrix with $\mathcal{P}=\sigma_x$ and $\mathcal{T}=*$ (i.e. complex conjugation) and its eigenvalues are given by $\lambda_{\pm}(k) =\pm\sqrt{\va^2 + \vb^2 + 2\va\vb\cos{k} - \gamma^2}$. Thus, the eigenvalues of $H_{PT}(k)$ go from being real at small $\gamma$ to complex-conjugate pairs when the gain-loss strength becomes large, i.e. $\gamma > \gamma_{PT}(k) \equiv \left|\va + \vb\E^{ik}\right|$. Therefore, the $\mathcal{PT}$-symmetry breaking threshold for $H_{PT}$, defined by Eqs.(\ref{eq:hssh}) and (\ref{eq:GammaPT}), is given by $\gamma_{PT} = |\va-\vb|$. We also note that for $\gamma>\vt \equiv (\va + \vb)$, all eigenvalues of the $\mathcal{PT}$-symmetric SSH Hamiltonian become complex. Therefore, in the following we will use $\vt$ as a convenient energy scale. The non-unitary time evolution operator $G(t)=\exp(-iHt)$ for each momentum $k$ can be written as
\begin{equation}
G_{PT}(k,t) =\cos\left[\lambda(k) t\right] - \I\frac{\vec{h}(k)\cdot\vec{\sigma}}{\lambda(k)}\sin\left[\lambda(k) t\right].
\end{equation}
We note that for $\gamma<\gamma_{PT}(k)$, the norm of $G_{PT}(k,t)$ oscillates with time and when $\gamma > \gamma_{PT}(k)$, it scales as $\exp[\pm2\Im\lambda(k) t]$ at long times. Thus after the $\mathcal{PT}$-breaking transition occurs, the intensity of an initially normalized state grows with time, leading to nonlinear effects that are not captured by the Hamiltonian $H_{PT}$.

The Hamiltonian for a lossy SSH model with $N=2M+1$ dimers (considered by RL) is given by $H_L=H_0+\Gamma_L$ where 
\begin{equation}
	\Gamma_L = -2\I\gamma\sum_{m=-M}^{M}\ket{m,B}\bra{m,B} = \Gamma_{PT} - \I\gamma{\bf 1}_N.
\end{equation}
An initially normalized state $|\phi(t)\rangle$ evolving under the lossy Hamiltonian $H_L$ decays with a rate given by $\partial_t\braket{\phi(t)|\phi(t)}=-4\gamma\sum_m |\braket{m,B|\phi(t)}|^2$. RL~\cite{Rudner2009} showed that the state's mean displacement $\langle\Delta m\rangle_L\equiv -4\gamma\sum_m\int_0^{\infty}dt\, m|\braket{m,B|\phi(t)}|^2$, a space-and-time integrated property, is equal to the winding number (0 or 1) of the lossy SSH Hamiltonian $H_L(k)$ in the momentum space. It is therefore robustly quantized. Because the $\PT$-symmetric and lossy dimer models are related by $H_{PT}= H_L +i\gamma{\bf 1}_N$, it is straightforward to define a similar mean displacement for a state $|\Psi(t)\rangle$ evolving under the $\mathcal{PT}$-symmetric Hamiltonian with both gain and loss, 
\begin{equation}
\braket{\Delta{m}} = -4\gamma\sum_{m=-M}^{M}\int_0^\infty dt \, e^{-2\gamma t}|\braket{m,B|\Psi(t)}|^2.
\label{eq:dm}
\end{equation}
It follows from RL that in an infinite $\mathcal{PT}$-symmetric lattice, an initial state $|\Psi(0)\rangle=|0,A\rangle$ will lead to $\langle\Delta m\rangle=0$ when $\va>\vb$ and $\langle\Delta m\rangle=1$ when $\va<\vb$. This transition is topological in origin and is therefore independent of the gain-loss strength $\gamma$, and insensitive to small perturbations of the Hamiltonian $H_{PT}$. 

Fig.~\ref{fig:overview} shows a summary of corresponding numerical results. The general initial state of the quantum walker confined to a single dimer can be written as $|\Psi(0)\rangle=|0,\psi_0\rangle$ where $\ket{\psi_0} = \cos(\theta/2)\ket{A} + e^{i\phi}\sin(\theta/2)\ket{B}$ is shown on the Bloch sphere in panel (a). The state at the north pole $|\psi_0\rangle=|A\rangle$ in the lossy dimer lattice was theoretically studied in RL and experimentally explored in Z, who also showed that when $|\psi_0\rangle=|B\rangle$, the south pole, the mean displacement is always zero. In the following sections we present results for the mean displacement $\langle\Delta m\rangle$, Eq.(\ref{eq:dm}), over the entire Bloch sphere. 

Figure~\ref{fig:overview}(b) shows the time-dependent intensity in each waveguide representing the dimer lattice, obtained from the time evolution $|\Psi(t)\rangle=G_{PT}(t)|\Psi(0)\rangle=\exp(-iH_{PT}t)|\Psi(0)\rangle$ of an initial state $|\Psi(0)\rangle=|0,A\rangle$. The lattice has $N=7$ dimers, intra-dimer coupling is $\va/\vt=0.75$ and the gain-loss strength $\gamma/\vt=0.5$.  In panel (c), we show the results for the mean displacement $\braket{\Delta m}$ for the two points on the Bloch sphere explored by RL and Z. When $\ket{\psi_0}=\ket{A}$ (blue), the mean displacement abruptly changes from one to zero when the intra-dimer coupling exceeds the inter-dimer coupling, i.e. $\va/\vt\geq 1/2$. The solid line represents results from a tight-binding calculation with $H_{PT}$ for a lattice with $N=41$ dimers and loss strength $\gamma=0.5\vt$. The open diamonds are results obtained from beam propagation method~\cite{Kawano2001, Wartak2013} for an array of $N=21$ dimers (see Refs.~\cite{Zeuner2015,Harter2016b,Harter2018} for details of waveguide parameters and calculations). When $\ket{\psi_0}=\ket{B}$, tight-binding calculations (red solid line) and BPM analysis (red open diamonds) show that the mean-displacement is zero over the entire range of $0\leq\va/\vt\leq 1$. Apart from the finite size effects that become important when the lattice size $N$ decreases, we see that the tight-binding and BPM results are consistent with each other, and with the analytical result that is applicable for an infinite lattice~\cite{Rudner2009}. These results for a $\mathcal{PT}$-symmetric lattice with $\langle\Delta m\rangle$ are, by construction, the same as those for the lossy dimer lattice with $\langle\Delta m\rangle_L$ investigated in RL and Z, and thus validate our analysis for a $\mathcal{PT}$-SSH model. The shaded area in panel (c) is the region where the $\PT$-symmetric dimer lattice is in the $\PT$-broken state, i.e. $|\va-\vb|\leq\gamma$. In this area, the net intensity experiences exponential amplification and therefore the nonlinear effects will become appreciable.

\section{Susceptibility of the topological transition to the initial state}
\label{sec:initial-state}

\begin{figure*}
\begin{center}
\includegraphics[width=\textwidth]{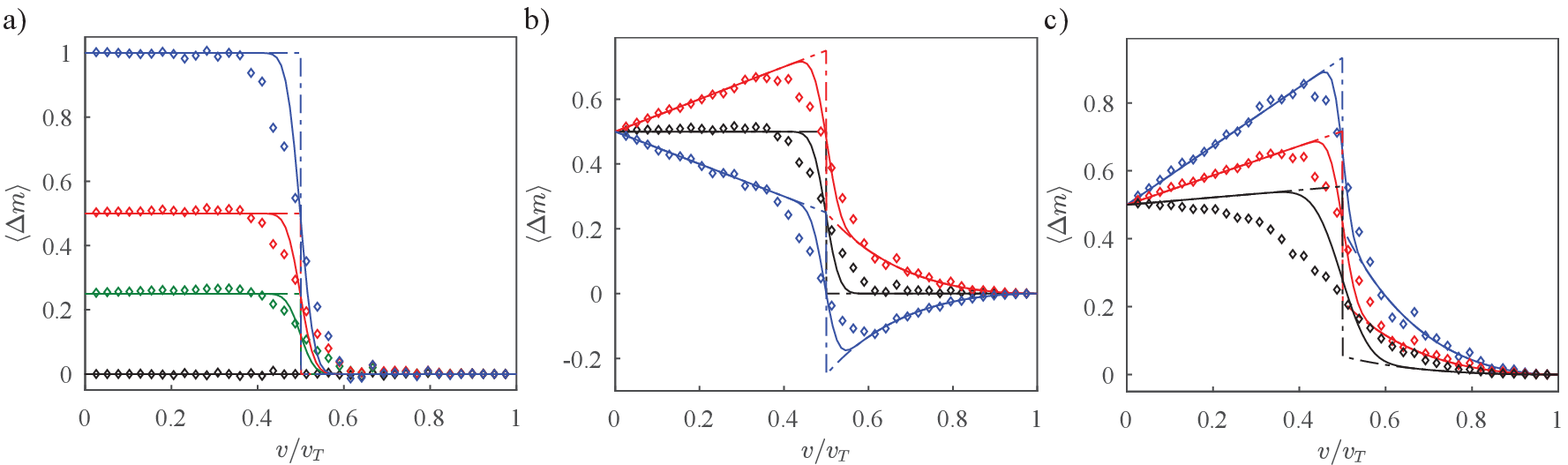}
\caption{\label{fig:dm_r} (a) Plot of $\braket{\Delta m}$ vs. $\va/\vt$ for $\theta=0$ (blue), $\theta=\pi/2$ (red), $\theta=2\pi/3$ (green) and $\theta=\pi$ (black), showing the reduction of the quantization value to $\cos^2\theta/2$. (b) Plot of $\braket{\Delta m}$ vs. $\va/\vt$ for $\theta=\pi/2$ with $\phi=0$ (black), $\phi=+\pi/2$ (red), and $\phi=-\pi/2$ (blue), showing the lack of quantization. (c) Plot of $\braket{\Delta m}$ vs. $\va/\vt$ for $\theta=\pi/2$ and $\phi=\pi/3$, with changing decay rates, $\gamma/\vt=0.25$ (blue), $\gamma/\vt=0.5$ (red), $\gamma/\vt=2$ (black), showing the gradual reduction of the non-quantized term with increasing $\gamma$. For both (a) and (b), the gain/loss rate is $\gamma/\vt = 0.5$, and in all figures (a-c), the dot-dashed lines indicate theoretical values, the solid lines indicate the tight binding results for a system of $N=41$ cells, and the diamonds indicate BPM results for $N=21$ cells. }
\end{center}
\end{figure*}

In this section, we consider the fate of the mean displacement $\braket{\Delta m}$, Eq.(\ref{eq:dm}) for an arbitrary initial state in the central dimer. The state $|\psi_0\rangle$ on the Bloch sphere, allows for different weights $\cos^2(\theta/2)$ and $\sin^2(\theta/2)$ on the gain and loss sites respectively. The relative phase difference $\phi\in[0,2\pi)$ between the $A$ and $B$ sites within a dimer imparts the state a nonzero transverse momentum. 

Figure~\ref{fig:dm_r} shows the mean displacement as a function of $\va/\vt$ for different values of $(\theta,\phi)$ and different gain-loss rates $\gamma$. Solid lines are from tight-binding calculations with an $N=41$ dimer lattice, and the open diamonds are from BPM simulations with $N=21$ dimers. The analytical results for an infinite lattice, which we will derive later in this section, are shown by dot-dashed lines. Panel (a) shows the behavior of $\langle\Delta m\rangle(\va/\vt)$ for real initial states ($\phi=0$) with $\theta=0$ (blue), $\theta=\pi/2$ (red), $\theta=2\pi/3$ (green) and $\theta=\pi$ (black). We see that the mean displacement changes abruptly from $\langle\Delta m\rangle=\cos^2(\theta/2)$ to $\langle\Delta m\rangle=0$ when the inter-dimer coupling exceeds the intra-dimer coupling at $\va/\vt=1/2$. Thus, for states confined to the red circle on the Bloch sphere, Fig.~\ref{fig:overview}(a), the mean displacement is still quantized, albeit between values of $\cos^2(\theta/2)$ and zero. These results are independent of the gain-loss strength $\gamma$.  

In Fig.~\ref{fig:dm_r}(b), for $\theta=\pi/2$, we show the result of changing the relative phase $\phi$ on the mean displacement $\langle\Delta m\rangle$. These results are for a gain-loss strength $\gamma/\vt=0.5$. When $\phi=\pi/2$ (red), the mean displacement first increases over its $\va/\vt=0$ value of $\cos^2(\theta/2)=0.5$ and then decreases around the transition point $\va/\vt=1/2$ returning to zero when $\va/\vt=1$. When $\phi=-\pi/2$ (blue), the mean displacement shows an opposite trend, becoming negative after the transition point and rising back to zero at $\va/\vt=1$. We remind the reader that the BPM simulations (open diamonds) are carried out on a waveguide array smaller in size than the tight-binding calculations (solid lines), and therefore, their deviation from the infinite-$\PT$-lattice analytical results (dot-dashed lines) is systematically larger. Panel (c) of Fig.~\ref{fig:dm_r} shows the dependence of $\langle\Delta m\rangle$ for an initial state with $(\theta,\phi)=(\pi/2,2\pi/3)$ on the strength $\gamma$ of the gain-loss potential. This dependence is manifest only when the initial state has a nonzero transverse momentum, $\phi\neq 0$. As the gain-loss strength increases from $\gamma/\vt=0.25$ (blue) to $\gamma/\vt=0.5$ (red) to $\gamma/\vt=2$ (black), the mean displacement is suppressed to its quantized value, i.e $\cos^2(\theta/2)=0.5$ or zero. 

The tight-binding and BPM results in Fig.~\ref{fig:dm_r} show that the mean displacement $\langle\Delta m\rangle$, Eq.(\ref{eq:dm}), is quantized and independent of $\gamma$ when the initial state is real, but that the topological quantization of $\langle\Delta m\rangle$ is modified with phase- and gain-loss strength-dependent contributions. In the following, we analytically derive both results. The mean displacement is the expectation value of the operator 
\begin{equation}
\op{Q}= -4\gamma\int_0^\infty dt e^{-2\gamma t} G_{PT}^\dagger(t)\sum_m |m,B\rangle m\langle m,B|G_{PT}(t),
\end{equation}
i.e. $\langle\Delta m\rangle=\langle\Psi_0|\op{Q}|\Psi_0\rangle$. Note that $\op{Q}$ is a global operator containing both spatial and temporal information. Converting the spatial-sum $\sum_m$ into Fourier space $k$-derivatives, and expressing the momentum-space time-evolved state on the B-sublattice as $g(k,t)\equiv\langle k,B|G_{PT}(k,t)|\Psi(0)\rangle$, the mean displacement can be written as
\begin{equation}
\braket{\Delta m}=-4\I\gamma\int_0^\infty dt\, e^{-2\gamma t}\oint\,\frac{dk}{2\pi}g^*(k,t)\partial_k g(k,t). 
\label{eq:dmint}
\end{equation}
To further simplify the integration terms in Eq.(\ref{eq:dmint}), we write $g(k,t) = -\I\cos(\theta/2) g_A(k,t)e^{i\Theta(k)} +\sin(\theta/2)e^{i\phi} g_B(k,t)$ where we have defined the following pair of real functions,
\begin{eqnarray}
	g_A(k,t) &=& u(k)\sin\left[\lambda(k) t\right]/\lambda(k), \\
	g_B(k,t) &=& \cos\left[\lambda(k) t\right] -\gamma\sin\left[\lambda(k) t\right]/\lambda(k), 
\end{eqnarray}
and the real quantities $u(k)$ and $\Theta(k)$ are the magnitude and phase of the complex number $(\va + \vb e^{i k})$. Integrating first in time and ignoring terms which are odd in $k$ or total derivatives under the integral $\oint dk$, we obtain
\begin{widetext}
\begin{equation}
\label{eqn:dmwinding}
\braket{\Delta m} =\cos^2(\theta/2)\oint\frac{dk}{2\pi} \partial_k\Theta(k)-4\gamma\sin\theta\sin\phi\oint \frac{dk}{2\pi}\sin\Theta(k)\int_0^\infty dt e^{-2\gamma t} g_B(k,t)\partial_kg_A(k,t). 
 \end{equation}
 \end{widetext}
The first integral is the winding number $W_\Theta$ of the function $\Theta(k)=\arg(\va+\vb e^{ik})$. We remind the reader that the winding number is zero for $\va>\vb$ and is one when $\va<\vb$. The second integral can be evaluated by using $\int_0^\infty dt e^{-2\gamma t} g_B(k,t)\partial_k g_A(k,t)=\partial_ku/8\gamma^2$. As a result, Eq.(\ref{eqn:dmwinding}) reduces to 
\begin{equation}
\label{eqn:dm-theory}
\braket{\Delta m}(\theta,\phi,\gamma)=\cos^2(\theta/2)W_\Theta+\frac{1}{4\mu}\sin\theta\sin\phi\frac{1}{\gamma},
\end{equation}
where $\mu^{-1} \equiv (1/2\pi)\oint dk(\partial_k u) \sin\Theta(k)$ can be further simplified to $\mu^{-1}=\min(\va,\vb^2/\va)$. 

Equation (\ref{eqn:dm-theory}) is the analytical result for the mean-displacement of a quantum walker in an infinite, $\mathcal{PT}$-symmetric lattice when the walker starts out in the central dimer, $m=0$, with an arbitrary initial state $|\psi_0\rangle$ on the Bloch sphere. The dot-dashed lines in Fig.~\ref{fig:dm_r} are obtained from Eq.(\ref{eqn:dm-theory}) and show that the key features of the topological transition $\va/\vt=1/2$ and its fragile aspects are both preserved in finite size systems (solid lines) with realistic parameters for BPM simulations \cite{Zeuner2015,Harter2016b,Harter2018} of the waveguide array (open diamonds). We see that $\braket{\Delta m}(\theta,\phi,\gamma)$ has a topological contribution from the winding number with a prefactor $\cos^2(\theta/2)$ as seen in Fig.~\ref{fig:dm_r}(a). It also has a non-topological contribution that is dependent on the initial state $(\theta,\phi)$, the gain-loss strength $\gamma$, and the dimer-lattice parameters $\va$ and $\vb$. This contribution has a quasi-classical interpretation: $p_0=\sin\theta\sin\phi$ is the dimensionless transverse momentum of the initial state $|\psi_0\rangle$, $\mu$ is the effective mass of the quantum walker, and this leads us to identify $1/4\gamma$ as the displacement time-scale $\tau$. With these definitions, Eq.(\ref{eqn:dm-theory}) becomes $\langle\Delta m\rangle=\langle\Delta m\rangle_{\phi=0}+(p_0/\mu)\tau$. Thus, the non-quantized contribution to $\langle\Delta m\rangle$ is just the classical drift of the quantum walker in time $\tau$. It is interesting to note that the effective mass of the walker $\mu$ is independent of all system parameters except the inter-dimer and intra-dimer couplings, and thus it can be used for the experimental determination of the coupling strengths. 

\begin{figure*}
\begin{center}
\includegraphics[width=\textwidth]{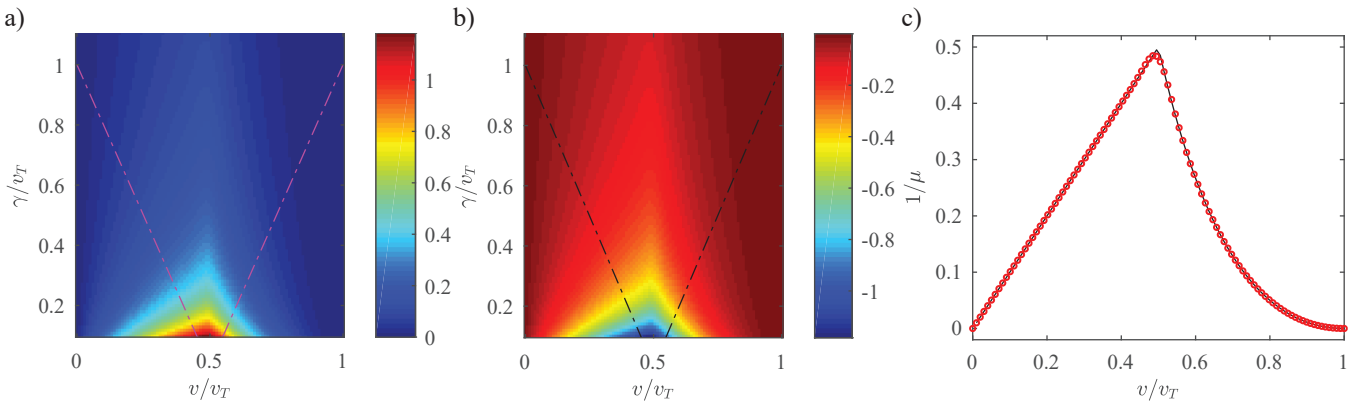}
\caption{\label{fig:dm_gamma} (a) Dependence of the quasiclassical contribution $\braket{\Delta m}-\braket{\Delta m}_{\phi=0}$ on the gain-loss strength $\gamma/\vt$ and the intra-dimer coupling $\va/\vt$. The initial state with $(\theta,\phi)=(\pi/2,\pi/2)$ has a positive transverse momentum, resulting in a positive quasiclassical contribution. This contribution is maximum in the vicinity of the topological transition threshold, $\va/\vt=1/2$, and decreases monotonically with $\gamma$. (b) When $\phi=-\pi/2<0$, the sign of the quasiclassical contribution is reversed. The dot-dashed line in both panels (a) and (b) represent the $\mathcal{PT}$-symmetric phase boundary. (c) The inverse effective mass $1/\mu$ as a function of $\va/\vt$ shows an excellent agreement between the theoretical and tight-binding results. The solid black line is the theoretical result for an infinite lattice, $\mu^{-1}=\min(\va,\vb^2/\va)$; open red circles are obtained by fitting the tight-binding results for a $\mathcal{PT}$-symmetric SSH lattice with $N=41$ dimers to the quasiclassical contribution formula, i.e. $\langle\Delta m\rangle-\langle\Delta m\rangle_{\phi=0}=(p_0/\mu)\tau$.}
\end{center}
\end{figure*}

Figure~\ref{fig:dm_gamma}(a) shows the dependence of this quasiclassical contribution $\braket{\Delta m}-\braket{\Delta m}_{\phi=0}$ on the intra-dimer coupling $\va/\vt$ and the gain-loss strength $\gamma/\vt$ for an initial state with $\theta=\pi/2$ and $\phi=+\pi/2$. We remind the reader that the topological contribution $\langle\Delta m\rangle_{\phi=0}$ is independent of the gain-loss strength. Since the initial state now has a positive momentum, the quasiclassical contribution is always positive. It is maximum at the smallest gain-loss strengths, vanishes when $\va=0$ or $\vb=0$, and is most dominant near the topological transition threshold $\va/\vt=1/2$. When $\phi=-\pi/2$, Fig.~\ref{fig:dm_gamma}(b), the momentum sign is changed and so is the sign of the quasiclassical contribution. In both panels, the dot-dashed lines represent the $\mathcal{PT}$-symmetry breaking phase boundary given by $\gamma=|\va-\vb|$. It is worth noting that the behavior of the quasiclassical contribution is smooth across this boundary and shows that the $\mathcal{PT}$ transition does not influence the topological transition. Panel (c) shows the dependence of the inverse effective mass, $\mu^{-1}$ on the intra-dimer tunneling $\va/\vt$. The solid line is the analytical result, $\mu^{-1}=\min(\va,\vb^2/\va)$ and the open red circles are results obtained by fitting the tight-binding results in Fig.~\ref{fig:dm_gamma}(a)-(b) and Fig.~\ref{fig:dm_r}(b)-(c) to the quasiclassical picture. 

In this section, we have shown that the topological transition in the mean-displacement of a quantum walker, discovered by RL~\cite{Rudner2009} and Z~\cite{Zeuner2015} for a lossy dimer lattice, is fragile with respect to the initial state of the walker on the Bloch sphere. In the general case, we have found that the mean displacement has a robust topological part and a quasiclassical contribution that depends on the system parameters. These results are valid for both the $\mathcal{PT}$-symmetric dimer lattices that we consider and the lossy dimer lattice considered by RL and Z. 

\section{Susceptibility of the topological transition to nonlinearity}\label{sec:nonlinearity}

\begin{figure*}
\begin{center}
\includegraphics[width=\textwidth]{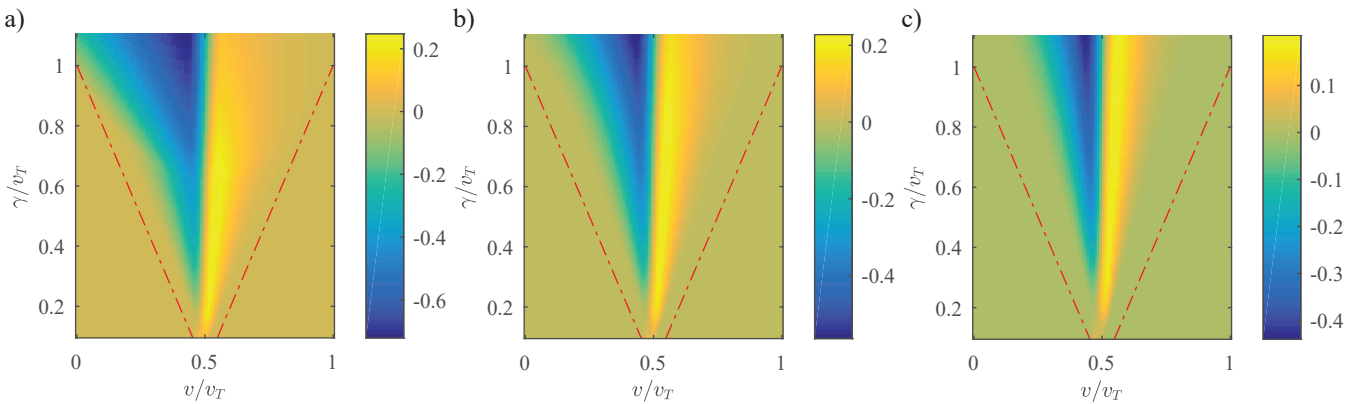}
\caption{\label{fig:dm_nonlinear} Change in the mean displacement, $\braket{\Delta m} -\braket{\Delta m}_{\eta=0}$, as a function of intra-dimer coupling and gain-loss strength for nonlinearities over three orders of magnitude: (a) $\eta/\vt=0.1$, (b) $\eta/\vt=0.01$, and (c) $\eta/\vt=10^{-4}$. The cone-shaped area between the red dot-dashed lines is the $\PT$-broken region, and the change in $\braket{\Delta m}$ is zero outside it. These results show that the topological transition in $\braket{\Delta m}$ survives in the $\mathcal{PT}$-symmetric phase, but is destroyed due to the nonlinear effects in the $\mathcal{PT}$-broken phase.}
\end{center}
\end{figure*}

A key difference between a $\mathcal{PT}$-dimer lattice with gain and loss, and a lossy dimer lattice (``passive'' $\PT$-system) is that the intensities in the former rapidly grow in the $\PT$-symmetry broken region \cite{Harter2016a}. In a waveguide-array realization of the $\PT$-symmetric SSH model, the large intensities change the local index-contrast in the waveguide due to the Kerr effect. In a Schr\"{o}dinger equation, it is encoded via a nonlinear, state-dependent potential that is dominant when the system is in the $\mathcal{PT}$-broken region. We remind the reader that the topological transition threshold $\va/\vt=1/2$ is always at the center of the $\mathcal{PT}$-symmetry broken region, Fig.~\ref{fig:overview}(c). In contrast, in the lossy dimer model, the intensities at all sites remain bounded, the net intensity of an injected pulse monotonically decreases with time, and the nonlinear effects are thus negligible. 

In this section, we investigate the influence of this nonlinearity on the topological transition. The Schr\"odinger equation for the system in the $\PT$-broken phase is given by 
\begin{equation}
\label{eq:nlse}
i\partial_t|\Psi(t)\rangle=H_{PT}|\Psi(t)\rangle+\eta\diag\left[|\Psi_m(t)|^2\right]|\Psi(t)\rangle,
\end{equation}
where $\eta$ is the Kerr cubic-nonlinearity coefficient and $|\Psi_m(t)|^2$ denotes the time-dependent intensity at site $m$. Generally, this discrete, nonlinear, $\PT$-symmetric Schr\"{o}dinger equation is not analytically soluble. Instead, we use tight-binding simulations for a $\PT$-lattice with $N=21$ dimers. Starting with initial state on the gain-site of the central dimer, i.e. $|\Psi(0)\rangle=|0,A\rangle$, we obtain the difference $\braket{\Delta m}-\braket{\Delta m}_{\eta=0}$ as a function of the intra-dimer coupling $\va/\vt$ and the gain-loss strength $\gamma$. For an infinite lattice with $\eta=0$, the mean displacement is just the winding number, i.e. $\braket{\Delta m}_{\eta=0}=W_\Theta=\{0,1\}$. However, to systematically account for the finite-size effects, we instead obtain both quantities numerically from the tight-binding model. It is worth emphasizing that to calculate $\braket{\Delta m}$ requires the time-dependent wavefunction $|\Psi(t)\rangle$, which, in turn, is obtained via time-ordered product due to the time-dependent effective Hamiltonian $H_{NL}=H_{PT}+\eta\diag|\Psi_m(t)|^2$. Since the on-site intensities $|\Psi_m(t)|^2$ grow exponentially with time in the $\PT$-broken region, this necessitates using increasingly smaller step-size for the time-ordered product calculations. 

Fig.~\ref{fig:dm_nonlinear} shows the change in the mean displacement as a function of strength of the nonlinearity. Recall that $\braket{\Delta m}_{\eta=0}=1$ for $\va/\vt< 1/2$ and $\braket{\Delta m}_{\eta=0}=0$ for $\va/\vt>1/2$. In each panel, the cone-shaped region between the two red dot-dashed lines denotes the parameter space in the $\va$-$\gamma$ plane where the $\PT$ dimer lattice is in the $\PT$-broken region. We see that $\braket{\Delta m}=\braket{\Delta m}_{\eta=0}$ when the system is in the $\PT$-symmetric phase. When $\eta/\vt=10^{-4}$, panel (c), $\braket{\Delta m}$ is suppressed from its unit value for weak intra-dimer coupling, and is increased from its zero value when the intra-dimer coupling exceeds the inter-dimer coupling. In contrast to the $\braket{\Delta m}_{\eta=0}$ result, the new mean displacement depends on the gain-loss strength and the change in $\braket{\Delta m}$ increases as $\gamma$ increases. As the strength of nonlinearity is increased from $\eta/\vt=10^{-2}$, in panel (b), to $\eta/\vt=10^{-1}$, in panel (a), the area where $\braket{\Delta m}$ changes become larger but is still confined to the $\PT$-symmetry broken region. We also see an asymmetry emerging between the regions of weak intra-dimer coupling and strong intra-dimer coupling. The results in Fig.~\ref{fig:dm_nonlinear} indicate that, in the $\PT$-symmetric regime, the topological quantization of the mean-displacement remains robust; however, inside the $\PT$-broken region (indicated by red dot-dashed lines), the nonlinearity dramatically affects this quantization. 


\section{Conclusion}
\label{sec:conclusion}

In this article, we have studied the fragile aspects of a topological transition in the mean displacement $\braket{\Delta m}$ that occurs in lossy or parity-time symmetric SSH models. We have shown that an arbitrary initial state of the quantum walker leads to $\braket{\Delta m}$ with a topological contribution, which is insensitive to the system parameters, and a quasiclassical contribution that strongly depends upon the system gain-loss strength and couplings. We have also shown that Kerr-effect generated nonlinearity affects the topological quantization of the mean displacement when the system is in the $\mathcal{PT}$-broken region, but preserves it in the $\mathcal{PT}$-symmetric region. 

Studying the instability of topological effects is a difficult problem because topological effects are, by definition, robust against small perturbations. We have shown that lossy and $\mathcal{PT}$-symmetric dimer lattices are amenable to such a study. The consistency among our analytical, tight-binding, and BPM simulation results implies that these fragile aspects, and in particular, the quasiclassical contribution to the mean displacement, can be observed in experiments on active and passive waveguide arrays.


\section*{Acknowledgments}
A.K.H. thanks the Los Alamos National Laboratory, where part of the work was carried out, for their hospitality. This work was supported in part by NSF Grant no. DMR-1054020 (Y.N.J.) and in part by the U.S. Department of Energy (A.S.). 


%

\end{document}